\documentclass[twocolumn,aps,epsfig,graphicx]{revtex4}
\usepackage[dvips]{graphics}
\newcommand{\be}{\begin{equation}}
\newcommand{\ee}{\end{equation}}

\newcommand{\bea}{\begin{eqnarray}}
\newcommand{\eea}{\end{eqnarray}}
\newcommand{\bd}{\begin{displaymath}}
\newcommand{\ed}{\end{displaymath}}
\newcommand{\bi}{\begin{itemize}}
\newcommand{\ei}{\end{itemize}}
\newcommand{\bc}{\begin{center}}
\newcommand{\ec}{\end{center}}
\newcommand{\bfl}{\begin{flushleft}}
\newcommand{\efl}{\end{flushleft}}
\newcommand{\bfr}{\begin{flushright}}
\newcommand{\efr}{\end{flushright}}

 \def\bq{{\bf q}}

\def\6{\partial}  
 \def\d{\delta}

\def\o{\omega}  \def\D{\Delta}

\def\={\!\!\!&=&\!\!\!}
\def\+{\!\!\!&&\!\!\!+~}
\def\-{\!\!\!&&\!\!\!-~}

\begin{document}
\title{Non-Fermi liquid behavior of the electrical resistivity at
the ferromagnetic quantum critical point}
\author{D. Bodea\cite{bordeaux}, M. Crisan, I. Grosu, and I. Tifrea\cite{iowa}}
\affiliation{Department of Theoretical Physics, University of
Cluj, 3400 Cluj-Napoca, Romania}

\begin{abstract}
We propose a model for the non-Fermi behavior in the proximity of
the quantum phase transition induced by the strong polarization of
the electrons due to local magnetic moments. The self-consistent
Renormalization Group methods have been used to calculate the
temperature dependence of the electrical resistivity and specific
heat. The $T^{5/3}$ dependence of resistivity and the $T \ln T$
dependence of the specific heat show that the magnetic impurities
drive a ferromagnetic quantum phase transition and near the
critical point the system present a non-Fermi liquid behavior. The
model is in good agreement with the experimental data obtained for
Ni$_x$Pd$_{1-x}$ alloy.
\end{abstract}

\maketitle

\section{Introduction}

Deviations from the standard Fermi liquid description in heavy
fermion systems (HFS) have been associated to the proximity of the
quantum phase transition (QPT). Defined as phase transitions at
$T=0$, QPT's are usually driven by quantum fluctuations controlled
by a non-thermal parameter, namely by impurities, pressure or
magnetic fields.\cite{1} Most of these low temperature magnetic
QPT's are from paramagnetic to antiferromagnetic state, but
recently HFS undergoing a paramagnetic to ferromagnetic phase
transitions were identified. Example of HFS which are close to
ferromagnetic order are Th$_{1-x}$U$_x$Cu$_2$Si$_2$,\cite{2}
MnSi,\cite{3} and Ni$_x$Pd$_{1-x}$.\cite{4} The itinerant-electron
ferromagnetic state in these materials is induced by pressure
(MnSi) or impurities (Th$_{1-x}$U$_x$Cu$_2$Si$_2$ and
Ni$_x$Pd$_{1-x}$).

Th$_{1-x}$U$_x$Cu$_2$Si$_2$ for concentrations $x\geq 0.15$
presents ferromagnetic order. Experimental data in these compounds
show a non-Fermi liquid behavior for $C/T\propto \ln{T}$ and a
critical temperature, $T_c$, as low as 12K for the $x=0.15$
sample.\cite{2} However, the magnetic susceptibility data present
a field dependance even at temperatures bellow $T_c$, a behavior
characteristic for weak itinerant ferromagnets.

In the case of MnSi experimental results showed the existence of a
first-order phase transition induced by pressure, the critical
temperature decreasing towards absolute zero at a pressure value
$p_c\simeq 14.6$ kbar. For $p>p_c$ a non-Fermi liquid behavior of
the resistivity as function of temperature is reported, $\rho\sim
T$.\cite{3} The phenomenological theory \cite{3} based on the
interaction of the electrons with the overdamped spin fluctuations
at low energies, can explain the occurrence of the non-Fermi
state. The quantum transport anomalies of the itinerant-electron
ferromagnetic state have been discussed also by Belitz,
Kirkpatrik, Narayanan and Vojta (BKNV) \cite{5}, the non-Fermi
behavior of the system being proved based on the scaling approach.
The model from Ref. \cite{5} predicts a first order phase
transition and it is in an excellent agreement with the QPT driven
by pressure in MnSi \cite{3}.

Ni$_x$Pd$_{1-x}$ at a Ni concentration $x_c\cong 0.025$ presents
ferromagnetic order assumed to occur in the itinerant electron
system of Pd atoms due to a strong polarization of these electrons
by magnetic impurities (in this case Ni). The transition critical
temperature depends on the Ni concentration as $(x-x_c)^{3/4}$ in
the critical region. In the same region of the phase diagram,
experimental data reveal a $T^{5/3}$ dependance of the relative
resistivity, $\D\rho$, and a $T\ln{T}$ dependance of the specific
heat, $C(T)$.\cite{4} All these experimental data clearly identify
the non-Fermi character of the electronic excitations close to the
quantum critical point (QCP). These results were successfully
described in terms of phenomenological spin-fluctuation
models.\cite{5bis}

In this work we extend the Doniah-Wohlfarth model,\cite{6}
proposed for the explanation of the itinerant electron
ferromagnetic state driven by impurities in Ni$_x$Pd$_{1-x}$
compounds, to the critical region and we calculate using
renormalization group methods the temperature dependence of the
resistivity and the specific heat. We will apply the Hertz-Millis
\cite{8,9} version of the Renormalization Group method (RNG) to
the Doniah-Wohlfarth model taking into consideration the quantum
effects at finite temperature and extract the effects of the spin
fluctuations on the system. The self-consistent renormalization
group theory, given by Moriya \cite{7} will be used to calculate
the temperature dependence of the electrical resistivity.

\section{Model}

We consider that the fluctuations in the magnetization in the critical
region are given by the action:
\begin{equation}
S_{eff} \; = \; S_{eff}^{(2)} \; + \; S_{eff}^{(4)}\;, \label{1}
\end{equation}
where
\begin{equation}
S_{eff}^{(2)} \; = \; \frac{1}{2} \sum_q \; \chi^{-1}(q) \; |\phi(q)|^2
\label{2}
\end{equation}
and
\begin{equation}
S_{eff}^{(4)} \; = \; \frac{u}{4} \sum_{q_1} \ldots \sum_{q_4} \;
\phi(q_1) \ldots \phi(q_4) \; \delta(q_1 + \cdots + q_4)\;.
\label{3}
\end{equation}
Here we introduce the notation $q = ({\bf q},\omega_n)$, $\omega_n$ being
the bosonic Matsubara frequencies, and
\[
\sum_q \; = \; k_B T \; \sum_n \; \int \frac{d^d {\bf q}}{(2
\pi)^d}\;.
\]
In Eqs. (\ref{2})-(\ref{3}) $\chi(q)$ is the fluctuation
propagator and $u$ is the coupling constant.

In the following we consider that the spin impurity has a very
strong polarization effect on the electrons and at a critical
concentration $x=x_c$ a new phase, which is in fact a
ferromagnetic phase, can be reached. This model has been proposed
first by Donniach and Wolhfarth \cite{6} using a single impurity
approximation. The susceptibility of the polarized electrons was
given (see Ref. \cite{6}) as
\begin{equation}
\chi({\bf q},\omega) \; = \; \frac{ \chi_0({\bf q},\omega) }{ 1 -
\left( I + 2\frac{J^2 R^{\prime}}{J R - \omega} \right)\;
\chi_0({\bf q},\omega)}\;, \label{4}
\end{equation}
where $\chi_0({\bf q},\omega)$ is the susceptibility of the
electronic system. In Eq. (\ref{4}) $J$ is the exchange
interaction between electrons and localized spins and $I$ is the
interaction between electrons. The parameters $R$ and $R^{\prime}$
have been calculated as
\begin{equation}
R \; = \; \frac{1}{N} \; \sum_k \left( n_{k \downarrow} - n_{k \uparrow}
\right) \label{5}
\end{equation}
and
\begin{equation}
R^{\prime}\; = \; - x \; \left< S^z \right>\;, \label{6}
\end{equation}
$x$ being the impurity concentration of the magnetic moments with spin ${\bf %
S}$.

The dynamic susceptibility $\chi _{0}({\bf q},\omega )$ has the form \cite{7}
\begin{equation}
\chi^{-1} _{0}({\bf q},\omega )\;\simeq \chi^{-1} _{0}(0,0)
\left[\;1\;-\;Dq^{2}\;+\;iC\frac{\omega }{q}\right]\;, \label{7}
\end{equation}
where $D$ and $C$ are constants. We approximate $\chi ({\bf q},\omega )$
from Eq. (\ref{4}) as
\begin{equation}
\chi ({\bf q},\omega )\;=\;\frac{1}{\chi _{0}^{-1}({\bf q},\omega )-I-2\frac{%
JR^{\prime }}{R}}\;, \label{8}
\end{equation}
a results which based on Eq. (\ref{4}) can be written as
\begin{equation}
\chi ({\bf q},\omega )\;=\;\frac{1}{\delta
_{0}(x)+aq^{2}-i\frac{\omega }{ \Gamma q}}\;, \label{9}
\end{equation}
where $a$ and $\Gamma$ are constants, and $\delta _{0}(x)$ is
given by
\begin{equation}
\delta _{0}(x)\;=\;\chi _{0}^{-1}\;-\;I\;-\;2\frac{JR^{\prime
}}{R}\;. \label{10}
\end{equation}
The parameter $\delta _{0}(x)$, with a linear dependance on the
impurity concentration, measures the distance from the QCP.

The model is valid only for systems in which the local moments
give a strongly polarization of the itinerant electrons and this
is typically for metals as Pd which is paramagnetic and the Stoner
criterium cannot be satisfied only due to the electron-electron
interaction. For this particular system we do not expect
localization effects, excepting the case of nonmagnetic
impurities, but this cannot drive the system in the ordered phase.

This model, described by Eqs. (\ref{1}), (\ref{2}) and (\ref{9}),
can be treated using the Renormalization Group method, in the
version proposed by Hertz \cite{8} and Millis \cite{9}. Following
Refs. \cite{8,9} we perform the standard scaling $k \rightarrow
k^{\prime}/b$, $\omega_n \rightarrow \omega_n^{\prime}/b^z$, $z$
being the dynamic critical exponent. We obtain the following flow
equations:\cite{11}
\begin{equation}
\frac{d T(l)}{d l} \; = \; z T(l)\;, \label{29}
\end{equation}
\begin{equation}
\frac{d \Gamma(l)}{d l} \; = \; (3-z) \Gamma(l)\;, \label{30}
\end{equation}
\begin{equation}
\frac{d \delta(l)}{d l} \; = \; 2 \delta(l) \; + \; 2(n+2) f_1
u(l)\;, \label{31}
\end{equation}
\begin{equation}
\frac{d u(l)}{d l} \; = \; [4-(d+z)] u(l) \; - \; (n+8)f_2 \;
u^2(l)\;, \label{32}
\end{equation}
where $d$ is the spatial dimension, $f_i$ are functions
characteristic for the model (see Ref. \cite{9}) and $l=\ln b$ is
the scaling variable. Additionally, the system free energy will
scale as:
\begin{equation}
\frac{d F(l)}{d l} \; = \; (d+z) F(l) \; + \; f_3\;, \label{34}
\end{equation}
$f_3$ being again a characteristic function of the model.

\section{Specific heat and resistivity}

The evaluation of the renormalized free energy $F[T(l)]$ based on
Eq. (\ref{34}) will lead to the temperature dependence of the
specific heat which by definition can be calculated as the second
derivative of the free energy with respect to the temperature,
$C(T)=-T(\6^2F/\6T^2)$. In general there will be two distinct
contributions to the renormalized free energy, associated to a
quantum domain, $T(l)\ll 1$, and to a classical one, $T(l)\gg 1$.
For more details on the calculation of the free energy see Refs.
\cite{9,11}. However, as the form of the system susceptibility
match the corresponding form for a ferromagnetic system, if we
consider the $d=3$ case, the specific heat is obtained as
\begin{equation}
C(T) \; = \; \gamma_0 T \; + \; \gamma_1 T \; \ln T\;, \label{38}
\end{equation}
$\gamma_0$ and $\gamma_1$ being constants, a result which clearly
show that the behavior of the considered system is non-Fermi, as
corrections to the linear temperature dependance of the specific
heat are logarithmic. This result is in agreement with the
experimental data presented in Ref. \cite{4}. We have to mention
that a similar behavior was obtained using RG method for a system
in the proximity of the Lifshitz quantum critical point.\cite{10}
Recent experimental data in silicon MOSFETs\cite{12} suggested a
QPT to a ferromagnetically ordered state in $d=2$. An analysis of
the specific heat behavior for the $d=2$ case was done in Ref.
\cite{11} suggesting a different temperature dependance of the
specific heat.

In order to calculate the temperature dependence of the
resistivity we apply the self-consistent theory of fluctuations to
the action given by Eq. (\ref{1}). This can be done in the version
of $1/n$ expansion ($n$ being the number of components of the
bosonic field $\phi$) applied to the $\phi^4$ action. Using the
approximation $|\phi^4| \sim 2 <|\phi^2|>|\phi^2|$ the
renormalized parameter $\d(x)$ can be calculated from the
following self-consistent equation:
\begin{widetext}
\begin{equation}
\delta(x) \; = \; \delta_0(x) \; + \; \frac{u}{2}
\left(\frac{n}{2}+1\right) \; k_B T \; \sum_{\omega_n} \; \int
\frac{d^3 {\bf k}}{(2 \pi)^3} \; \frac{1}{\delta(x) + k^2 +
\frac{|\omega_n|}{\Gamma k}}\;. \label{11}
\end{equation}
The summation over the bosonic Matsubare frequencies on the second
term in the right hand side (rhs) of Eq. (\ref{11}) can be
performed analytically leading to the following expression:
\begin{equation}
\delta(x) \; = \; \delta_0(x) \; + \; \frac{u}{2} \left(\frac{n}{2}+1\right) \; \int \frac{%
d^3 {\bf k}}{(2 \pi)^3} \; \int_0^{\Gamma k} d \omega \; \coth \left( \frac{%
\omega}{2 k_B T} \right) \frac{\frac{\omega}{\Gamma k}}{(\delta +
k^2)^2 + \left( \frac{\omega}{\Gamma k} \right)^2}\;. \label{12}
\end{equation}
The temperature dependence of the QPT parameter $\delta(x)$ can be
extracted if we consider Eq. (\ref{12}) at the QCP in order to
eliminate the bare QPT parameter $\d_0(x)$. Accordingly, we obtain
\begin{eqnarray}
\delta(x) & = & \frac{u}{2} \left(\frac{n}{2}+1\right) \; \int
\frac{d^3 {\bf k}}{(2 \pi)^3} \; \int_0^{\Gamma k} \frac{d
\omega}{\pi} \left[ \coth \left(\frac{\omega}{2 k_B T}\right) -1
\right] \; \frac{\frac{\omega}{\Gamma k}}{(\delta(x) + k^2)^2 +
\left( \frac{\omega}{\Gamma k} \right)^2} \nonumber \\
& &+  \frac{u}{2} \left(\frac{n}{2}+1\right) \; \int \frac{d^3
{\bf k}}{(2 \pi)^3} \; \int_0^{\Gamma k} \frac{d \omega}{\pi}
\left[ \frac{\frac{\omega}{\Gamma k} }{(\delta(x) + k^2)^2 +
\left( \frac{\omega}{\Gamma k} \right)^2} -
\frac{\frac{\omega}{\Gamma k}}{k^4 + \left( \frac{\omega}{\Gamma
k} \right)^2} \right]\;. \label{14}
\end{eqnarray}
\end{widetext}
For $T=0$ we will show that $\delta \ll T$ holds. In this
approximation the first integral in the rhs of Eq. (\ref{14})
becomes:
\begin{equation}
I_1 \; \cong \; 2 k_B T \; \int \frac{d^3 {\bf k}}{(2 \pi)^3} \;
\int_0^{k_B T} \frac{d \omega}{\omega} \frac{\frac{\omega}{\Gamma
k }}{A^2 + \left( \frac{ \omega}{\Gamma k} \right)^2}\;,
\label{15}
\end{equation}
where $A = \delta(x) + k^2$. Performing the integral over $\omega$
we get:
\bea
I_1 \; &=& \; \frac{\Gamma}{6 \pi^3} \; \left( \frac{k_B
T}{\Gamma} \right)^{4/3} \; \int_0^{\infty} dy \; \frac{\arctan
y}{y^{4/3}}\nonumber\\&=&C_1 \; (k_B T)^{4/3}\;, \label{16}
\eea
where $y = k_B T / \Gamma k^3$; $y_c$, associated with the upper
critical wave-vector $k_c$, has been substituted by infinity.
$C_1$ is a constant. The second integral from the rhs of Eq.
(\ref{14}), $I_2$, is:
\be
I_2= \int_0^{\Gamma k} \frac{d \omega}{2 \pi} \left[
\frac{\frac{\omega}{\Gamma k }}{A^2 + \left( \frac{\omega}{\Gamma
k} \right)^2} - \frac{\frac{\omega}{\Gamma k}}{k^4 + \left(
\frac{\omega}{\Gamma k} \right)^2} \right]\;, \label{18}
\ee
and can be performed if one introduce a new variable $y = k /
\delta^{1/2}$ with an upper cut-off $y_c = k_c
[\delta(x)]^{-1/2}$. The final result can be express as:
\begin{equation}
I_2 \; = \; - C_2 \; \delta(x)\;. \label{20}
\end{equation}
Based on Eqs. (\ref{14}), (\ref{16}) and (\ref{20}) we obtain for
$\delta(x)$ the following temperature dependance:
\begin{equation}
\delta(x,T) \; = \; C(u,n) \; (k_B T)^{4/3}\;, \label{21}
\end{equation}
which is finite for $n \rightarrow \infty$. This result can be
inverted in order to extract the concentration dependence of the
critical temperature. If one consider that the temperature can be
approximated by the critical value ($T_c$), and that in the first
order $\delta(x)\approx \delta_0(x)$ we have
\be
T_c(x)\sim (x-x_c)^{3/4}\;,
\ee
a result which is in agreement with the one discussed in Ref.
\cite{4}.

The temperature dependance of the resistivity can be extracted as
the imaginary part of the self-energy, obtained as a result of
electrons interacting with the ferromagnetic fluctuations. In the
one-loop approximation we have:
\begin{equation}
\Sigma ({\bf k},i\omega _{n})\;=\;g^{2}k_{B}T\;\sum_{{\bf
q},i\omega _{l}}G( {\bf k}+{\bf q};i\omega _{n}+i\omega
_{l})\;D({\bf q},i\omega _{l})\;, \label{22}
\end{equation}
where $g^{2}$ is the coupling constant, $G({\bf k},i\omega _{n})$ is the
electronic Green function,
\begin{equation}
G({\bf k},i\omega _{n})\;=\;\frac{1}{i\omega _{n}-\epsilon ({\bf
k})}\;, \label{23}
\end{equation}
with $\omega _{n}=(2n+1)k_{B}T$, and $D({\bf q},i\omega _{l})$ has
the same form with $\chi(\bq,\o)$ given by Eq. (\ref{9}) whit
$\delta_{0}(x)$ replaced by $\delta(x)$. Performing the summation
over $\omega _{l}$ in Eq. (\ref{22}) we obtain:
\begin{eqnarray}
&&\Sigma ({\bf k},i\omega _{n})=-\pi g^{2}\;\sum_{{\bf q}
}\int_{-\Gamma q}^{\Gamma q}\frac{dz}{\pi }\frac{\frac{z}{\Gamma
q}}{A^{2}+\left( \frac{z}{\Gamma q}\right) ^{2}} \nonumber \\
&&\times[n_{B}(z)+n_{F}(z+\omega )]\;\delta \lbrack \omega
+z-\epsilon ( {\bf k}+{\bf q})]\;, \label{24}
\end{eqnarray}
where $n_{B}(z)$ is the Bose function and $n_{F}(z)$ is the Fermi function.
For $|z|\ll T$ we approximate $n_{B}(z)+n_{F}(z)\sim T/z$ and performing the
analytical continuation $i\omega _{n}\rightarrow \omega +i\eta $ we
calculate
\bea
&&-Im\;\Sigma ^{R}(k_{F},0)\;\simeq \;\nonumber\\
&&\pi g^{2}\;\sum_{{\bf q} }\int_{-\Gamma q}^{\Gamma
q}\frac{dz}{\pi }\frac{\frac{T}{\Gamma q}}{ A^{2}+\left(
\frac{z}{\Gamma q}\right) ^{2}}\;\delta \lbrack z-\epsilon ( {\bf
k}+{\bf q})] \;\;\;\;\;\;\label{25}
\eea
which leads to the following expression for the imaginary part of
the self-energy
\begin{equation}
-Im\;\Sigma
^{R}(k_{F},0)\;=\;\text{const}\frac{(k_{B}T)^{3}}{\delta (T)}\;.
\label{26}
\end{equation}
Using now Eq. (\ref{21}) we obtain the temperature dependence of the
scattering time $1/\tau _{eff}=-{Im}\Sigma ^{R}$ as:
\begin{equation}
\frac{1}{\tau _{eff}}\;=\;\text{const}\;(k_{B}T)^{5/3}\;,
\label{27}
\end{equation}
which gives for the temperature dependent resistivity $\Delta \rho
(T)=\rho (T)-\rho (0)$ the following behavior
\begin{equation}
\Delta \rho (T)\;\sim \;(k_{B}T)^{5/3}\;, \label{28}
\end{equation}
in agreement with the experimental data from Ref. \cite{4}.

\section{Conclusion}

In conclusion, we presented a model for the ferromagnetic QPT
driven by the magnetic impurities which polarize the Fermi liquid
close to the Stoner instability. The result can be regarded as a
generalization of the Doniach and Wohlfarth \cite{6} mean-field
model for the case of QPT. As a general result we conclude that
the system presents a non-Fermi behavior in the critical region
around a QCP. The non-Fermi character of the system is sustained
by a $T\ln{T}$ behavior of the specific heat correction term, and
by an electrical resistivity which presents a $T^{5/3}$
dependence. Both these results are in good agreement with the
experimental data reported in Ref. \cite{4}. A similar temperature
dependance of the resistivity was obtained by Mathon\cite{mathon}
using a simple molecular field theory. However, despite a good
agreement between the calculated and experimental values of the
resistivity, Mathon calculations are not able to explain the
non-Fermi behavior of the specific heat and do not take into
account the quantum effects observed in Ni$_x$Pd$_{1-x}$, making
the model inadequate for the proper description of the QPT in this
particular system. A phenomenological description of the QPT was
also made by Lonzarich.\cite{5bis} We also explain the
concentration dependance of the critical temperature, our result
being in good agreement with the experimental data.

We mention that this model is different from the BKNV \cite{5}
model, in which the spin susceptibility for the $d=3$ case is
considered as $\chi(q, \omega=0) = \chi_0 + q^2 \ln (p_F/q)$. This
form of $\chi(q)$ has been carefully analyzed by Millis \cite{13}
and at the present time new experimental data are needed for a
confirmation of this spin susceptibility. Recently, Belitz and
Kirkpatrick \cite{14} have been reconsidered the QPT in the clean
itinerant-electron ferromagnet. The coupling of the order
parameter fluctuations to the soft fermionic fluctuations lead to
a theory which is very different than theories based on the
Hertz-Millis model. The main point in the new version of the BKNV
theory is that the fluctuations can change the first order phase
transition in a second order one. However, the occurrence of two
dynamical exponents $z$ and $\tilde{z}$ for the two kind of
fluctuations makes the two theories very different, even if in any
case the mean field behavior can explain the experimental data.

\end{document}